\documentclass[namedreferences]{SolarPhysics}
\usepackage[optionalrh]{spr-sola-addons} 
\usepackage{graphicx}                    
\usepackage{color}                       
\usepackage{url}                         

\newcommand{\adv}{    {\it Adv. Space Res.}}

\newcommand{\aap}{    {\it Astron. Astrophys.}}

\newcommand{\apj}{    {\it Astrophys. J.}}

\newcommand{\jgr}{    {\it J. Geophys. Res.}}

\newcommand{\solphys}{  {\it Solar Phys.}}

\newcommand{\ssr}{    {\it Space Sci. Rev.}}

\begin{document}

\begin{article}

\begin{opening}

\title{Stationary stagnation point flows in the vicinity of a 2D magnetic null point}
\subtitle{I. Systems with vanishing electric field and an X-type magnetic null point}

%

\author{D.H.~Nickeler\sep 
        M.~Karlick\'{y}\sep
	M.~B\'{a}rta
	}


\institute{Astronomical Institute, AV \v{C}R, Fri\v{c}ova 298,
           25165 Ond\v{r}ejov, Czech Republic  email: \url{nickeler@asu.cas.cz}\\ 
	   email: \url{karlicky@asu.cas.cz} email: \url{barta@asu.cas.cz}\\
	   }
			
\begin{abstract}
The appearance of eruptive space plasma processes, e.g., in eruptive flares as
observed in the solar atmosphere, is usually assumed to be caused
by magnetic reconnection. The process of magnetic reconnection is often
connected with singular points of the magnetic field. 
We therefore analyse the system
of stationary resistive/non-ideal magnetohydrodynamics (MHD)
in the vicinity of singular points of flow
and field to determine the boundary between
reconnection solutions and non-reconnective solutions.
We find conditions to enable the plasma to cross
the magnetic separatrices also inside the current sheet, close to the
current maximum. The results provide us with the topological and
geometrical skeleton of the resistive MHD fields. We therefore have to
perform a local analysis of almost all non-ideal MHD solutions without
a specific non-idealness.
We use Taylor expansions of the magnetic field, the velocity field
and all other physical quantities, including the non-idealness,
and with the method of
a comparison of the coefficients, the non-linear resistive MHD system is solved
analytically.
In the vicinity of a stagnation point, it is reasonable to assume
that the density is constant.
We find that the electric field has to be zero and that the
non-ideal term/resistivity has to depend on the spatial coordinates and cannot be constant,
otherwise it has to be zero everywhere. It turns out that not every non-ideal
flow is a \lq reconnective\rq~flow and that pure resistive/non-ideal MHD only allows for
such reconnection-like solutions, even if the non-idealness is localized to
the region around the magnetic null point.
It is necessary that the flow close
to the magnetic X-point is also of X-point type to guarantee
positive dissipation of energy and annihilation of magnetic flux. 
If the non-idealness has only a one-dimensional, sheet-like structure, only one
separatrix line can be crossed by the plasma flow, similar to reconnective
annihilation solutions.
\end{abstract}
			
			
\end{opening}

\section{Introduction}

Magnetic reconnection is thought to be a process, being responsible
for many eruptive plasma phenomena in space plasmas and astrophysical plasmas,
like geomagnetical substorms or eruptive flares.
Although magnetic reconnection in two dimensions (2D) is fairly well understood, e.g., see the
comments in \inlinecite{baty}, it would be interesting to have more detailed
informations about the topological and geometrical structure of flow and field
lines in the vicinity of the singular points of plasma flow and magnetic field.

Classical reconnection scenarios, see \inlinecite{petschek}
or Sweet-Parker model (see, e.g., \opencite{sweet}) propose a magnetic null point
and a stagnation point flow into the diffusion region, i.e. the stagnation point
is inside this diffusion region.
 
\inlinecite{priest75} where the first to analyse the case of incompressible
2D MHD with constant resistivity.
It turned out that either the magnetic field must be of higher order in the
spatial variables $x$ and $y$ or that there is no stagnation point flow, but
a shear flow. Therefore their result is, that the classical \lq hyperbolic\rq~
stagnation point flow needs higher order terms, concerning
the spatial variables ($x$ and $y$). 
Solutions locally containing only higher order terms do not
allow for topologically/structurally stable magnetic fields.
Therefore either no reconnection can take place, as in the case of higher order
terms no \lq hyperbolic\rq~magnetic field can exist, or the stagnation point
flow is not of hyperbolic type. 

Annihilation solutions have been studied, where
\inlinecite{craighenton} chose a special ansatz for the solution
of the resistive MHD to get reconnection solutions.
They emphasize the first order momentum equation and neglect the
energy transport (equation) or rather the entropy conservation (equation), starting
with a nonlinear perturbation of magnetic annihilation solutions. This lead them
to so called \lq reconnective annihilation\rq~solutions, where only one of the two
separatrix-lines are crossed, and the other is only tangent to the
converging streamlines. The current sheet has a one-dimensional
structure (straight line). The results found by \citeauthor{craighenton}, and
\inlinecite{craigrickard} confirm the results found
earlier by \inlinecite{priest75}, who found more \lq shear-like\rq~
flows instead of typical stagnation point flows.

Later on \inlinecite{tassi} and \inlinecite{titov2} extended the method
to curvelinear current sheets.

It was shown by \inlinecite{priest94}
and later on in extended form by \inlinecite{watsoncraig} that under
certain circumstances (like constant resistivity or current depending/anomalous 
resistivity and sub-Alfv\'enic flow etc.) reconnection is
impossible, the so called anti-reconnection theorems.

In 3D for constant resistivity a careful analysis of topologically different
solutions has been discussed in \inlinecite{titov1}. In 2D such an analysis
is missing and should be done here, but without taking the restriction to constant
resistivity into account.

In contrast to the aforementioned models we do not search for special solutions,
but the most {\it general} solution without constant or special non-constant
resistivity or a specific non-idealness. Thus our analysis covers all forms of
non-ideal terms. In the sense of topological fluid dynamics
we want to get insight how the field and streamlines are rooted in the null point(s)
of flow and field, i.e. which geometrical shapes of field and streamlines are possible
and which not. We define and investigate the influence of an \lq effective\rq~non-constant 
resistivity and use hereby
the full energy equation of resistive MHD instead of using only the assumption
of incompressibility.

The problems concerning (exact) analytical models very often are:
\begin{itemize}

\item The physical quantity \lq resistivity\rq~is often only a constant smallness parameters

\item if really gradients of the resistivity $\eta$ and maybe also the amplitude/absolute
value are recognized as important for the non-ideal process, what are
the \lq shapes\rq~of such resistivities enabling magnetic reconnection?

\item it is not clear which {\it flow} topologies are allowed to generate
reconnection, reconnective annihilation solutions, or other solutions
for general non-ideal or resistive terms

\end{itemize}


Thus our aims are: 
\begin{itemize}
\item To show not every non-ideal or resistive process in the vicinity of a null point
is a magnetic reconnection process

\item it is not enough to have a localized resistivity to get a reconnective
solution, finding parameters that mark the boundary between reconnection and
non-reconnection solutions

\item why should there be no reconnection process  close to the
magnetic null point and what happens if we do not restrict
ourselves to complete incompressible dynamics without energy transport?

\item finding analytical {\it and} exact solutions close to the null point

\item detailed investigation of topological skeleton of flow and magnetic field in the frame
of resistive MHD; it is an analysis of the resistive system close to singular points of
flow and magnetic field

\item performing linearization, necessary to get information
about the skeleton of magnetic reconnection

\end{itemize}

We will concentrate on the topology and geometrical properties of flow and
field and discuss different cases with respect to their implications for
magnetic reconnection.

To summarize:
The main goal is to get the topological and geometrical skeleton of the resistive MHD
equations, i.e., if the Jacobian of the magnetic field at the magnetic null point
has only real Eigenvalues (hyperbolic null point), which Eigenvalues can be found for the
Jacobian of the plasma velocity at the stagnation point? In our analysis the null point
of the velocity field (=stagnation point), should have an almost identical position and 
a vanishingly small offset to the magnetic null point. The resistvity $\eta$ should be
a positive quadratic form in $x$ and $y$ close to the magnetic null point,
indicating that the resistivity indicates
real dissipation in the form of Ohmic heating and annihilation of magnetic flux,
see also the explanations and calculations in the section \ref{sws}.
How does the geometrical structure of the velocity field look like,
and which kind of Eigenvalues are allowed by the resistive or in general
non-ideal MHD equations?

\section{Assumptions and basic equations}

\subsection{The topological and geometrical structure of the magnetic field}
The topological classification of 2D vector fields in the vicinity
of their null points is described in the textbooks (with connection about phase
portraits of dynamical systems, i.e. systems of ordinary differential equations)
e.g. of \inlinecite{arnold}$\!\!,$\inlinecite{amann}$\!$ or \inlinecite{reitmann}$\!\!$.
The topological structure of magnetic fields in the vicinity of null points is
described, e.g., by\inlinecite{parnell}$\!\!$ and concerning the construction of ideal MHD
flows, e.g. in\inlinecite{nigo}$\!\!$.

Our interest is to ask which topology and geometry of the
{\it macroscopic} flow correspond to which magnetic topology and geometry in the
frame of MHD. In contrast to the analyses mentioned in the previous paragraph
we here have to investigate the topological and geometrical properties of
{\it both} vector fields, i.e. that of the plasma flow and that of the magnetic
field. To perform this analysis we solve the resistive MHD system with a linear
ansatz for both vector fields, i.e., we perform a Taylor expansion of both vector
fields in the vicinity of their null points, neglecting derivatives higher than
the first ones.

That implies that we allow the stagnation point (= null point of the velocity field)
of the velocity field $\vec{\rm v}$ to have a small offset concerning the
magnetic null point, but neglecting here, the influence of the second
derivatives of the velocity field. An influence would be noticeable
for significant offsets, but is negligible for very small offsets, justifying
our assumptions. We will justify this assumptions afterwards at the end of
section \ref{clf}, estimating the influence of an offset caused by the
gravitational force in the case of completely linear velocity and magnetic fields,
where the second derivatives of the vector fields vanish automatically.

Now let $x_{0}$ be the offset in $x$-direction and $y_{0}$ the offset
in $y$-direction.
Then we can express the class of linear velocity fields with the help of their
first derivatives, i.e. Jacobians:
\begin{equation}
\vec{\rm v}= \stackrel{\leftrightarrow}{\rm\bf V} \vec x =
\left(
\begin{array}{cc}
V_{11} & V_{12} \\
V_{21} & V_{22} \\
\end{array}
\right)
\left(
\begin{array}{c}
x-x_{0}\\
y-y_{0}\\
\end{array}
\right)
\label{2}\end{equation}
in analogy to the magnetic field $\vec B$
\begin{equation}
\vec{\rm B}= \stackrel{\leftrightarrow}{\rm\bf B} \vec x =
\left(
\begin{array}{cc}
B_{11} & B_{12} \\
B_{21} & B_{22} \\
\end{array}
\right)
\left(
\begin{array}{c}
x\\
y\\
\end{array}
\right)
\label{2b}\end{equation}

\begin{figure}[h!]
\resizebox{\hsize}{!}
{\includegraphics{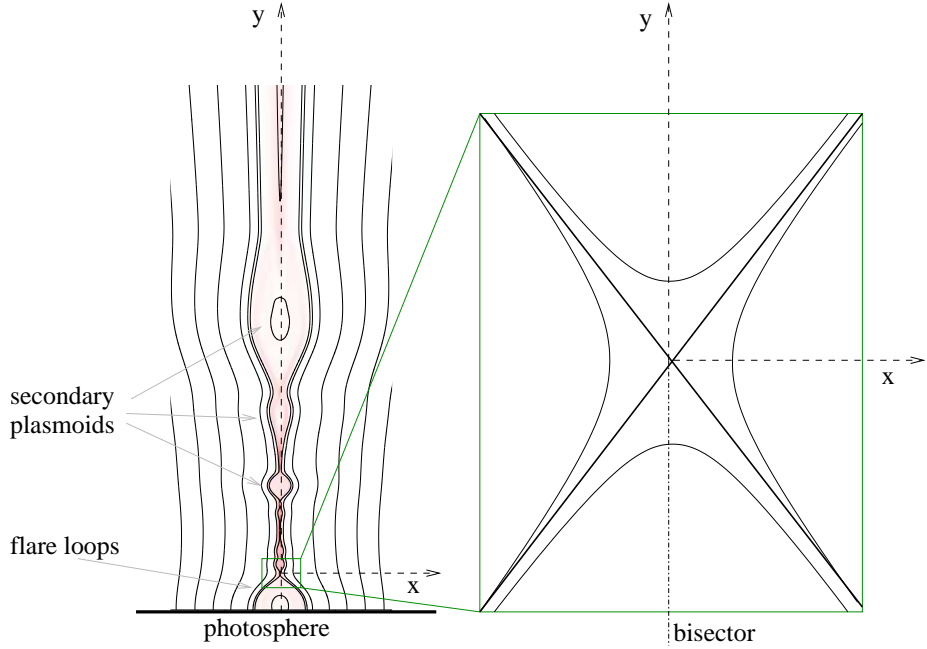}}
\caption{Scenario of a flare loop with an X-point.}
\label{bild0}
\end{figure}

According to\inlinecite{parnell}$\!\!$, every magnetic field of the form
of Eq.\,(\ref{2b}) can be
rotated such, that it is represented by a magnetic
flux function\footnote{This form of the magnetic flux function
implies that the bisector of the separatrix angle should be approximately
perpendicular to the photosphere, i.e. we assume the bisector of the separatrix
angle to be symmetric with respect to the solar surface.} $A=ax^2+by^2$,
so that a standard null point of the magnetic field
appears with a constant current density $j_{z}=j_{0}=$ const around the origin, where
the magnetic field is given by $\vec\nabla A\times\vec e_{z}=\vec B$.
The corresponding scenario of a flare loop, with its symmetry
axis being almost perpendicular to the photosphere, is given in Fig.\ref{bild0}.

The current density is
\begin{equation}
\vec\nabla\times\vec B=-\Delta A\, \vec e_{z}=\mu_{0} j_{z}\,\vec e_{z}\, ,
\label{1}\end{equation}
such that
\begin{equation}
\vec{\rm B}=
\left(
\begin{array}{cc}
B_{11} & B_{12} \\
B_{21} & B_{22} \\
\end{array}
\right)
\left(
\begin{array}{c}
x\\
y\\
\end{array}
\right)
\equiv
\left(
\begin{array}{cc}
0 & 2b\\
-2a & 0\\
\end{array}
\right)
\left(
\begin{array}{c}
x\\
y\\
\end{array}
\right)\, .
\\ \label{1b}\end{equation}
The different topologies of 2D vector fields
are represented by two independent parameters, the
threshold current $j_{t}$, and the Eigenvalues of $\stackrel{\leftrightarrow}{B}$,
the $\lambda_{B}$'s, or $j_{z}$, the current in $z$--direction
\begin{equation}
\lambda_{B}=\pm\frac{\mu_{0}}{2}\sqrt{j_{t}^2 - j_{z}^2}\, ,
\label{1c}
\end{equation}
implying a bijective relation between the variables
$a$ and $b$ and the threshold $j_{t}$ and the actual current $j_{z}$, to
be precise
\begin{eqnarray}
&& a=-\frac{\mu_{0}}{4}\left( j_{t} + j_{z}\right)\quad
b=\frac{\mu_{0}}{4}\left( j_{t} - j_{z}\right)
\\
\Leftrightarrow && j_{z}=-\frac{2}{\mu_{0}}\left( a + b \right)\quad
j_{t}=\frac{2}{\mu_{0}}\left( b - a \right)\, .
\end{eqnarray}
We will use the four variables $a,b,j_{t},j_{z}$
if it is convenient, i.e. combinations of the
four to simplify the corresponding terms in the equations.

The Eigenvalue $\lambda_{B}$ determines the topological structure
of the field and the geometrical shape of the field lines.
For a divergence free case there are only three main types of such fields:

\begin{itemize}

\item
the case that the Eigenvalue is zero ($|j_{t}|=|j_{z}|$)
corresponds to the one-dimensional current sheet
(degenerated case).

\item the case that $|j_{t}|<|j_{z}|$ corresponds to field lines
being topological circles (geometrical ellipses), if $j_{t}=0$ ($a=b$), then
geometrical circles. All the cases mentioned in the last sentence are so
called O-points.

\item the case of the so called X-points,
where $|j_{t}|>|j_{z}|$.

\end{itemize}
Only for the case in the last item a magnetic separatrices exist.
Such separating field lines must exist to enable magnetic reconnection.
A necessary condition for magnetic reconnection in 2D is that the plasma
flow crosses magnetic separatrices, see,
e.g.,\inlinecite{priestforbes}$\!\!$,\inlinecite{vasyliunas}$\!\!$,\inlinecite{cowley}$\!\!$,
\inlinecite{sonnerup1}$\!\!$, or\inlinecite{sonnerup2}$\!\!$.
The current free case is given by $j_{z}=0$, i.e. $a=-b$ and should be
excluded, as dissipation in such a case would have nothing to do with the
electric current.

\subsection{The basic resistive MHD equations and assumptions}

The following analysis is restricted to {\it pure} resistive dynamics, i.e.
Ohmic heating/dissipation without any other loss terms like viscosity
or heat conduction.
We choose the coordinate system in such a way, that the gravity is directed in 
negative $y$-direction (the unit vector in $y$-direction is $\vec e_{y}$).
The basic resistive stationary MHD equations in 2D are given by (following, e.g.,
\opencite{goedbloed})
\begin{eqnarray}
\vec\nabla\cdot(\rho \vec{\rm v}) &=& 0\, ,\label{4}\\
\rho(\vec{\rm v}\cdot\vec\nabla)\vec{\rm v} &=& \vec j\times\vec B - \vec\nabla p
- \rho g \vec e_{y}\,\, , \, g>0\label{5}\\
E_{0}+{\rm v}_{x} B_{y}- {\rm v}_{y} B_{x}&=& \eta j_{z}\label{6}\, ,\\
\gamma p \vec\nabla\cdot\vec{\rm v} +
\vec{\rm v}\cdot\vec\nabla p &=& (\gamma-1)\eta j_{z}^2 
\label{7}\, ,\\
\vec\nabla\times\vec B &=& \mu_{0}\vec j\label{7b},\\
\vec\nabla\cdot\vec B &=& 0\label{7c}\, .
\end{eqnarray}
Due to stationary Maxwell equations the electric field in the 2D case has
to be constant, i.e. $E_{z}\equiv E_{0}=$const see, e.g., \inlinecite{titov2}.
The same shape of the equations occur, if we introduce a generalized non-ideal term
on the right-hand side of Ohm's law. As the current density is constant close
to the X-point (see Eq.\,\ref{1b}), the shape of an effective resistivity
is that of the non-idealness
$\vec R$ (where only the $z$-component is non-zero), writing
\begin{eqnarray}
\vec\nabla\cdot(\rho \vec{\rm v}) &=& 0\, ,\\
\rho(\vec{\rm v}\cdot\vec\nabla)\vec{\rm v} &=& \vec j\times\vec B - \vec\nabla p
- \rho g \vec e_{y}\,\, , \, g>0\\
E_{0}+{\rm v}_{x} B_{y}- {\rm v}_{y} B_{x}&=& \vec R\cdot\vec e_{z}=R_{z}\, ,\\
\gamma p \vec\nabla\cdot\vec{\rm v} +
\vec{\rm v}\cdot\vec\nabla p &=& (\gamma-1) R_{z} j_{z} 
\, ,\\
\vec\nabla\times\vec B &=&\mu_{0}\vec j ,\\
\vec\nabla\cdot\vec B &=& 0\, .
\end{eqnarray}
The implication is that $\eta=\eta_{\rm effective}\propto R_{z}$, as $j_{z}\neq 0$
is constant close to the null point. 
The non-idealness resulting from two-fluid approach can be written for the
one fluid approach, e.g., following\inlinecite{braginsky}$\!\!x$, as
\begin{equation}
\vec R=\eta\vec j+\frac{1}{n e}\left(\vec j\times\vec B
- \vec\nabla\cdot\stackrel{\leftrightarrow}{p_{e}}\right) + \frac{m_{e}}{n e^2}
\displaystyle\left(
\displaystyle\langle\frac{\partial\vec j}{\partial t}\rangle_{t} + \vec\nabla\cdot\left(\vec{\rm
 v}\vec j + \vec j\vec{\rm v}
\right)\right)\, ,
\end{equation}
where $n$ is the particle density, $e$ the absolute value of the electron charge,
$\vec j\times\vec B$ the Hall term,
$\stackrel{\leftrightarrow}{p_{e}}$ the anisotropic electron pressure tensor,
$\frac{\partial\vec j}{\partial t}$ the electron acceleration term (in our case averaged
over a \lq fitting\rq~ characteristic time scale, as our approach is quasi-stationary), and
the tensor divergence $\vec\nabla\cdot\left(\vec{\rm v}\vec j +
\vec j\vec{\rm v}\right)$
describes the influence of electron inertia.
But even this two-fluid model does not include all terms which may be
(additionally) responsible
for \lq anomalous\rq~resistivity or non-idealness, like, e.g., electron scattering
by wave turbulence generated due to the Buneman 
instability (often considered, e.g., \opencite{buechner} and \opencite{karlicky}),
ionization-recombination terms, non-linear electron-ion momentum transfer terms,
electron and ion viscosities, radiative losses/gains, and so on.
As we do not specify the non-idealness, having
introduced the effective resistivity,
our analysis can be used for all physical models with different types
of non-ideal terms. For a more specific view but for non-local models of reconnection
with a generalized Ohm's law, see, e.g.,\inlinecite{craigwatson}$\!\!$.
The term $R$ is therefore only a {\it necessary}
\lq trigger\rq~enabling reconnection, not a sufficient one, see, e.g.,
\inlinecite{ks}$\!\!$.

We will calculate the (almost) complete solution space of the resistive
(non-ideal) MHD close to the null points of flow field and magnetic field.
The aim is to find the general correlation between the
Jacobians of the plasma velocity and the magnetic field, and the shape of the
unspecified non-idealness. 
At first both vector fields are treated as completely linear,
i.e. unbounded fields in an unbounded domain, see section \ref{clf}.
This does not exclude the possibility that in a bounded domain the fields at
the boundary are maybe not analytic, but the extrapolation of the field may
produce reasonable finite and regular solutions.

In section \ref{coli} we concentrate on the linearized fields
and take only first order terms of the spatial variables into account.
A similar method to determine the structure of the non-ideal term and the
flow and the magnetic field has been proposed and done, but for the case $j_z=0$
and for \lq global\rq~fields, in the frame of a toy model in\inlinecite{nifa}$\!\!$.

With this linear or linearized fields we can draw conclusions with respect
to the other MHD quantities, like pressure, density and resistivity: 
The lowest order of the magnetic field is linear, the current density is constant,
and therefore the Lorentz force $\vec j\times\vec B$ is linear in $x$ and $y$. This
is the reason that all other terms should be also at (the) most of first order in the spatial
variables.
This leads to the conclusion that the plasma pressure $p$ is at highest order quadratic
in $x$ and $y$, to allow for a \lq linear\rq~pressure force, i.e. we can express $p$ by
\begin{eqnarray}
p=p_{0}+p_{1} x + p_{2} y + p_{3} xy + p_{4} x^2 +p_{5} y^2\, ,
\label{8}\end{eqnarray}
where $p_{0}$ to $p_{5}$ are constant coefficients.

 As the velocity close to the stagnation point can be represented as a
linear term in $x$ and $y$,
the $\vec {\rm v}\cdot\vec\nabla \vec{\rm v}$ term is also linear in $x$ and
$y$, or constant.
The mass continuity equation, Eq.\,(\ref{4}) implies therefore
that the mass density, $\rho$, has to be constant in lowest order close to
the stagnation point.

Of course $\rho$ could be a simple linear function close to the null point
but that would imply
that the mass density has to vanish on a surface, or rather straight line in
2D, depending on the gradient of the density. This null-line {\it could}
be somewhere inside the \lq linear\rq~region.
In one direction the density could
then become negative. We assume that the density $\rho$ is constant
in the vicinity of the null points of magnetic and flow field to prevent
$\rho$ to have a negative or zero value within the domain of interest.
In addition it is a reasonable assumption that close to a stagnation point there
is a \lq stagnation region\rq~, i.e. the density has a maximum or minimum close
to the stagnation point.

In the case of completely linear fields it is now evident that the mass
continuity equation requires a divergence free velocity field, if we assume
all terms to have the same or comparable orders in $x$ and $y$. 

One has also to take into consideration that at the magnetic null point $E_{0}$
equals $\eta j_{z}$ and if we assume that the resistivity is constant, then
$E_{0}=\eta j_{z}$ everywhere in the vicinity of the neutral point. 
The same holds for special cases of non-constant resistivity,
e.g., if $\eta$ is a function of the current density $j_{z}$ only, one can
conclude that in the case of a linear field or in the vicinity of the null
point the current density is constant and therefore also the function
$\eta=$const, as $\eta=\eta(j_{z}=const)=$const .

The electric field $E_{0}$ is constant also outside the linear region, but the
resistivity and current density are, of course, not constant there.
This implies that the term ${\rm v}_{x} B_{y} -{\rm v}_{y} B_{x} $ equals to zero
within the linear region and therefore the flow is field-aligned everywhere
{\it within this region} in lowest order of the spatial variables.
To get not such \lq trivial\rq~solutions, which are non-reconnective in lowest
order,
we have to regard the resistive Ohm's law as a {\it definition equation} for the
spatially dependent resistivity, i.e. $\eta=\eta (x,y)$. Therefore, to enable
reconnection in the vicinity of the null point, the resistivity cannot be a
function of the current density only. We designate and regard this coefficient
as an \lq effective\rq~resistivity or short resistivity, even if this coefficient
originates from collisional theory. As the current density is constant inside
within the region of the linear field approximation, the (effective) resistivity
is a substitute expression for a general non-ideal term, determining, but vice
versa also determined by, the flow and magnetic field line structure! This
relation between the velocity or flow field $\vec{\rm v}$, the
magnetic field $\vec B$ and the resistivity $\eta$ will be analysed
in this article.

One can also recognize that Eq.\,(\ref{7}) (together with the incompressibility)
at the stagnation point would require
a vanishing resistivity in the absence of additional dissipation terms.
Written with all coefficients and comparing with all orders of $x$ and $y$ we get the
following system of equations, first from the Euler or momentum equation
\begin{eqnarray}
\rho\left(V_{11}^2 + V_{12} V_{21}\right) + 2 p_{4} - 2 a j_{z} &=& 0\, ,\label{9}\\
\rho\left(V_{11}^2 + V_{12} V_{21}\right) + 2 p_{5} - 2 b j_{z} &=& 0\, , \label{10}\\
p_{3} &=& 0\, , \label{11}\\
\rho (V_{21} - V_{21})V_{11} x_{0}- \rho( V_{12} V_{21} +
V_{11}^2) y_{0} + p_{2} +\rho g &=& 0\, ,\label{12} \\
-\rho V_{11}^2 x_{0}-\rho V_{12} V_{21} x_{0} + p_{1} &=& 0\, ,
\label{13}\end{eqnarray}
and the equations from the energy equation, first neglecting source or
heating terms, only including Ohmic heating,
\begin{eqnarray}
 (-V_{11} p_{1} -V_{21} p_{2})x_{0} + (V_{11} p_{2} -V_{12} p_{1})y_{0} &=&
q E_{0}\, , \label{14}\\
 V_{21} p_{2} +V_{11} \left( p_{1} -2 x_{0} p_{4}\right) - 2V_{12}y_{0} p_{4} &=&
2aq( V_{11} x_{0} + V_{12}y_{0}) \label{15}\\
 V_{12} p_{1} - 2 V_{21} x_{0} p_{5} + V_{11}\left( 2  y_{0} p_{5} - p_{2}\right) &=&
2bq\left(  V_{21} x_{0} - V_{11} y_{0}\right) \label{16} \\
 2 V_{12} p_{4} + 2 V_{21} p_{5} &=& q(-2a V_{12} - 2b V_{21})\, ,\label{17}\\
 2p_{4} V_{11} &=& -2a q V_{11} \, ,\label{18} \\
 -2p_{5} V_{11} &=& 2b q V_{11}\, , \label{19}
\end{eqnarray}
where $q:=(\gamma-1) j_{z}$.
The equations above are ordered after their physical meaning, the first five equations
Eqs.\,(\ref{9}) - (\ref{13}) correspond to the first order momentum equations,
while Eqs.\,(\ref{14}) - (\ref{19}) represent the terms
of the energy equation.  The Eqs.\,(\ref{12}), (\ref{13}) and (\ref{14}) are of zeroth
order in $x$ and $y$.
The Eqs.\,(\ref{9}), (\ref{10}), (\ref{15}) and (\ref{16}) are of first order in $x$
and $y$, the Eq.\,(\ref{17}) is of the order $xy$ and Eq.\,(\ref{18})
of second order in $x$, respectively Eqs.\,(\ref{19}) is of second order in $y$.

\section{Results}

In the first subsection we concentrate on linear fields, i.e. fields that
are unbounded in an unbounded domain. In the second subsection we take only
zero and first order terms of $x$ and $y$ into account, i.e.
we regard a Taylor expansion of maximum order one.

\subsection{Completely linear fields}\label{clf}

If $V_{11}\neq 0$ then from Eqs.(\ref{18}) and (\ref{19}) it follows
\begin{eqnarray}
p_{4} &=& -qa \,\label{20}\\
p_{5} &=& -qb\label{21}\, ,
\end{eqnarray}
and with two terms of the momentum equation, namely Eq.\,(\ref{9}) and
Eq.\,(\ref{10}),
excluding the current free case $j_{z}=0$, i.e. $a=-b$ and
the case with O-points, i.e. $a=b$, we infer that
\begin{eqnarray}
\rho(V_{11}^2 + V_{12} V_{21})=2qa + 2a j_{z}\label{22}\, , \\
\rho(V_{11}^2 + V_{12} V_{21})=2qb + 2b j_{z}
\label{23}\, , 
\end{eqnarray}
thus $q=(\gamma-1) j_{z}=- j_{z}$ should be valid, leading to $\gamma=0$.
As $\gamma=0$ is not representing a realistic equation of state, $V_{11}$ must be zero,
{\it taking terms of every order into acount}.

This is unaffected by the neglecting other source terms, not mentioned here,
as the interesting heat conduction term is of the order zero in $x$ and $y$,
and therefore affecting the Eq.\,(\ref{14}) only.

Now we use the assumption that $V_{11}=0$. Then we rewrite the MHD equations:
\begin{eqnarray}
\rho V_{12} V_{21} + 2 p_{4} - 2 a j_{z} &=& 0\, , \nonumber \\
\rho V_{12} V_{21} + 2 p_{5} - 2 b j_{z} &=& 0\, , \nonumber \\
-\rho V_{12} V_{21} y_{0} + p_{2} +\rho g &=& 0\, , \nonumber \\
-\rho V_{12} V_{21} x_{0} + p_{1} &=& 0\, , \nonumber\\
\nonumber\\
- V_{12} y_{0} p_{1} -  V_{21} x_{0} p_{2} &=& q E_{0}\, , \nonumber\\
V_{21} p_{2}- 2 V_{12} y_{0} p_{4} &=& 2 q V_{12} a y_{0}\, , \nonumber\\
V_{12} p_{1}- 2 V_{21} x_{0} p_{5} &=& 2 q V_{21} b x_{0}\, , \nonumber\\
V_{12} \left( p_{4} + q a\right) + V_{21} \left( p_{5} + q b\right) &=& 0 \, .
\label{24}\end{eqnarray}

With Eqs.\,(\ref{6}) and (\ref{7}) it can clearly be recognized that
at the stagnation point the resistivity must be zero. As the electric field
$E_{0}$ is constant and $E_{0}=\eta j_{z}$ at the magnetic null point
and the stagnation point, we conclude that the electric field
is zero (everywhere). The effect of a vanishing compressibility is in resistive
MHD also connected with a vanishing electric field.

In addition, as the $x$-offset $x_{0}$ in the fourth equation
(a zeroth order term of the force) of the system (\ref{24}) couples
an additional pressure gradient in $x$-direction to this offset only, this
offset can be set to zero without loss of generality. Therefore, from $x_{0}=0$
and $E_{0}=0$ we infer $p_{1}=0$, see the fourth and fifth equation of the system (\ref{24})
(part of the energy/entropy equation).

If we would further allow $y_{0}=0$ in the case of non-vanishing gravity,
then there would exist only static solutions either (as in this case
$V_{21}$ must be zero it implies that $V_{12}=0$), or 
the other possible solution branch would need an
adiabatic exponent $\gamma=0$.

For the non-degenerated case ($y_{0}\neq 0$) the general solution for the system
(\ref{24}) has two branches:
\begin{eqnarray}
p_{2} &=& -\frac{\rho g}{2} - \frac{\rho ^2 g V_{12}^2}{4 b \gamma j_{z}}
\mp \frac{\rho g}{4 b \gamma j_{z}}\,\sqrt{D}\, ,\nonumber\\
p_{4} &=& \frac{1}{2}\left( \left( 2 a - b\gamma \right) j_{z} +
\frac{\rho V_{12}^2}{2} \pm \frac{1}{2}\,\sqrt{D}\right)\,\nonumber \\
p_{5} &=& \frac{1}{2} \left( \left( 2-\gamma\right)b j_{z} +
\frac{\rho V_{12}^2}{2} \pm \frac{1}{2} \sqrt{D}\right)\, , \nonumber\\
y_{0} &=& \frac{\rho g}{2 b \gamma j_{z}} \, , \nonumber\\
V_{21} &=& \frac{2 \gamma b j_{z} - \rho V_{12}^2 \mp \sqrt{D}}{2 \rho V_{12}} \, ,
\label{25}
\end{eqnarray}
where the discriminant $D$ must fulfill
\begin{eqnarray}
D=8 a \rho V_{12}^2 \gamma j_{z} + \left( 2 b \gamma j_{z} - \rho V_{12}^2 \right)^2 \geq 0
\, 
\label{26}\end{eqnarray}
to guarantee that $V_{21}$ is real.
This leads to
\begin{eqnarray}
&& D=\left(\rho V_{12}^2 \right)^2 + \left( 8a -4b\right)\gamma j_{z} \rho V_{12}^2
+ 4 b^2\gamma^2 j_{z}^2 \geq 0 \label{27} \\
\Rightarrow && D_{2}:=\left[\left(4a-2b\right)\gamma j_{z} \right]^2
- 4 b^2 \gamma^2 j_{z}^2 \geq 0
\, , \label{28}
\end{eqnarray}
where $D_{2}$ is the discriminant of the quadratic equation Eq.\,(\ref{27}).
The inequality Eq.\,(\ref{28}) guarantees a real value
of $\rho V_{12}^2$. 
That leads together with the convention $j_{t}>0$
\begin{eqnarray}
\left( a-b\right) a \leq 0\quad\Leftrightarrow\quad j_{z}\geq -j_{t} \, .
\label{29}
\end{eqnarray}
This implies that either the actual current density
$j_{z}\geq 0$, then the field lines are hyperbolas or topological circles,
or $j_{z}< 0$, then only hyperbolic field lines or (anti-)parallel are possible.

Let us now estimate the influence of the offset in $y$-direction on the solution,
e.g., for the situation in the corona.
We can express the offset in $y$-direction by using the equation for the
offset $y_{0}$ in Eq.\,(\ref{25})
\begin{equation}
y_{0}=\frac{6\rho g}{5\mu_{0}\left(j_{t}/j_{z}-1\right)j_{z}^2}\, , 
\end{equation}
assuming that $\gamma=5/3$. 

We demand that the configuration should really be an X-point, but not
a one-dimensional sheet, therefore we assume that the
relation between the two characteristic currents is given by a value not
much larger than $j_{z}/j_{t}\approx 0.999$ (correponds to an opening
angle of the smaller separatrix angle of only about 2 degrees).
With the electron charge $e$, the number density $n$ and the
drift velocity ${\rm v}_{d}$ and $j_{z}=n e {\rm v}_{d}$,
it follows for $y_{0}$
\begin{equation}
y_{0}=1200\,\frac{\rho g}{\mu_{0} j_{z}^2}=
1.2\times 10^{3}\,\frac{n m_{p} g}{\mu_{0} n^{2} e^{2} {\rm v}_{d}^{2}}
\end{equation}
The value $|y_{0}|$ must now be compared with the typical lengthscale
and values of a magnetic field close to a null point, i.e. the coronal
magnetic field is about $B_{c}\approx 10^{-2} T$ (the index \lq c\rq~stands
for \lq coronal\rq) and the scale on which the
magnetic field varies linearly should be larger than a Debye length
(for coronal parameters in the range of mm to cm)~\footnote{
$\lambda_{D}=\left( ne^2/\left(\varepsilon_{0}k_{B}T\right)\right)^{1/2}
\approx 10^{-3}$m},
but much smaller than the active region loops ($l_{c}\approx 10^{7.5}$m)
to get an upper limit for $|y_{0}|$. We therefore use the approximate
current sheet thickness $d$. To estimate the current sheet thickness of
a finite current sheet, we assume again that the current density
$j_{z}= ne{\rm v}_{d}$, where ${\rm v}_{d}$ is the drift velocity, 
which should not exceed the thermal velocity (of the electrons),
i.e. ${\rm v}_{d}\leq\sqrt{3 k_{B} T/m_{e}}$. 
Let $k_{B}=1.38\times 10^{-23}$JK$^{-1}$ the Boltzmann constant,
$\varepsilon_{0}=8.85\times 10^{-12}$A$^{2}$s$^{4}$kg$^{-1}$m$^{-3}$
the dielectric constant, $m_p=1.67\times 10^{-27}$kg
the mass of the proton, $m_e=9.1\times 10^{-31}$kg
the mass of the electron, $n=1.6\times 10^{14}$m$^{3}$ the number density,
and $T_{d}=10^{6}$K (as the thermal temperature is about
$T=2\times 10^{6}$K, we consider an effective temperature $T_{d}$, thus that the
corresponding drift velocity ${\rm v}_{d}$ is definitely
below the critical drift velocity given by temperature $
{\rm v}_{th}^2=3 k_{B} T/m_{e}$
to guarantee a stable stationary current, instead of an kinetic instability
see, e.g., \opencite{papadopoulos}) the approximate temperature
(for the coronal parameters see, e.g., \opencite{stix}).
We choose a \lq small\rq~current density to get the upper limit of $y_{0}$.
The offset is then given by
\begin{equation}
|y_{0}|=\frac{1.2\times 10^{3} m_{p} g}{\mu_{0} n e^{2} {\rm v}_{d}^{2}}
=\frac{1.2\times 10^{3}\, m_{p} g}{\mu_{0} n e^{2} \left(\!\sqrt{3 k_{B} T_{d}/m_{e}}\right)^{2}}
\approx 3\times 10^{-6}{\rm m}\, .
\end{equation}
The lower bound of the current sheet thickness is approximately given by
\begin{equation}
d\approx \frac{B_{c}}{\mu_{0} j_{z}}=
\frac{B_{c}}{\mu_{0} n e \sqrt{3 k_{B} T_{d}/m_{e}}}\approx 33{\rm m}\, .
\end{equation}
The fraction of both values should be $y_{0}/d\ll 1$and is then determined by
\begin{equation}
\frac{y_{0}}{d}=1.2\times 10^{3}\,\frac{m_{p} g}{e {\rm v}_{d} B_{c}}
=\frac{1.2\times 10^{3}\, m_{p} g}{e \sqrt{3 k_{B} T_{d}/m_{e}} B_{c}}\approx
6\times 10^{-8}\ll 1 \, .
\end{equation}
Thus for typical coronal parameters the offset in $y$-direction
has a value of about mm or even some orders of magnitude less, in contrast to
the value of the typical lengthscale of the linear X-point region which is at
least of the order of one meter or some tenth of meters or even more. This is
an argument or rough justification to neglect the term of the interaction
between the shift and the second derivative of $\vec{\rm v}$.

Even if we assume that the drift velocity ${\rm v}_{d}$ is, let us say six orders,
smaller than the thermal velocity, the fraction $|y_{0}|/d$ would be of the order
of some percent. Therefore, even in this extreme regime of the drift velocity,
the stagnation point is located within the current sheet, assuming that
the sheet is really two-dimensional (non-singular) and its extensions in
both coordinate directions are about of the same order.
The complete mentioned discussion is valid exactly only in the case of unbounded
fields, but it shows that in general the offset due to gravity can be neglected.

\subsubsection{Discussion of the discriminant $D$}

The discriminant $D_{2}$ in Eq.\,(\ref{28}), being larger than zero is only a
{\it necessary} criterion. 
That $\rho V_{12}^2 > 0$ has also to be guaranteed. 

To fullfil the relation $D\geq 0$ we define two values, namely
\begin{eqnarray}
 \epsilon_{1}:=\rho V_{12, crit1}^2 &=& \frac{\mu_{0}}{2}\left( 3 j_{t} + j_{z}\right)
\gamma j_{z} + 2 \mu_{0}\gamma |j_{z}| \sqrt{\frac{ j_{t}
\left( j_{t} + j_{z}\right)}{2}}\\ 
\epsilon_{2}:=\rho V_{12, crit2}^2 &=& \frac{\mu_{0}}{2}\left( 3 j_{t} + j_{z}\right)
\gamma j_{z} - 2 \mu_{0}\gamma |j_{z}| \sqrt{\frac{ j_{t}
\left( j_{t} + j_{z}\right)}{2}}\\
\nonumber\label{30}
\end{eqnarray}
to represent the relation Eq.\,(\ref{27}) 
\begin{eqnarray}
&& \left(\rho V_{12}^2 - \epsilon_{1}\right) \left(\rho V_{12}^2 - \epsilon_{2} \right)
\geq 0\\
\Rightarrow\quad
\textrm{C(ase)I} && \rho V_{12}^2\geq \epsilon_{1} \quad\land\quad  \rho V_{12}^2\geq \epsilon_{2}
\label{31}\\
\textrm{is valid or} && \nonumber\\
\Rightarrow\quad
\textrm{C(ase)II} && \rho V_{12}^2\leq \epsilon_{1} \quad\land\quad  \rho V_{12}^2\leq \epsilon_{2}
\label{32}\, ,\end{eqnarray}
which implies for CI $\rho V_{12}^2 \geq {\rm max}(\epsilon_{1}, \epsilon_{2})
=\epsilon_{1}$ and for CII $\rho V_{12}^2\leq {\rm min}(\epsilon_{1}, \epsilon_{2})=
\epsilon_{2}$. The values $\epsilon_{1,2}$ are explicit functions of the current
densities $j_{t}$ and $j_{z}$. This gives us restrictions for $V_{12}$:
for $j_{z}\geq 0$ the lower bound for $V_{12}$ is given by
the left relation of Eq.\,(\ref{31}) otherwise the lower bound is zero,
writing
\begin{eqnarray}
\rho V_{12}^2 &\geq & \frac{\mu_{0}}{2}\left( 3 j_{t} + j_{z}\right)
\gamma j_{z} + 2 \mu_{0}\gamma |j_{z}| \sqrt{\frac{ j_{t}
\left( j_{t} + j_{z}\right)}{2}} \, ,
\label{33}\end{eqnarray}
such that the non-admissible region is then given by
\begin{eqnarray}
V_{12} \not\in \left[+V_{12,crit1} , -V_{12,crit1}\right]\, ,
\end{eqnarray}
and $V_{12,crit1}$ is given by
\begin{eqnarray}
V_{12,crit1}=\sqrt{\frac{\frac{\mu_{0}}{2}\left( 3 j_{t} + j_{z}\right)
\gamma j_{z} + 2 \mu_{0}\gamma |j_{z}| \sqrt{\frac{ j_{t}
\left( j_{t} + j_{z}\right)}{2}}}{\rho}}\, .
\label{34}\end{eqnarray}
For $j_{z}\leq 0$ all values of $V_{12}$ are allowed. There are no restrictions
for $V_{12}$, as for $j_{z}=0$ the critical value $V_{12,crit1}=0$ and for $j_{z}<0$ all
values of $V_{12,crit1}$ are imaginary.

\begin{figure}[h!]
\resizebox{\hsize}{!}
{\includegraphics{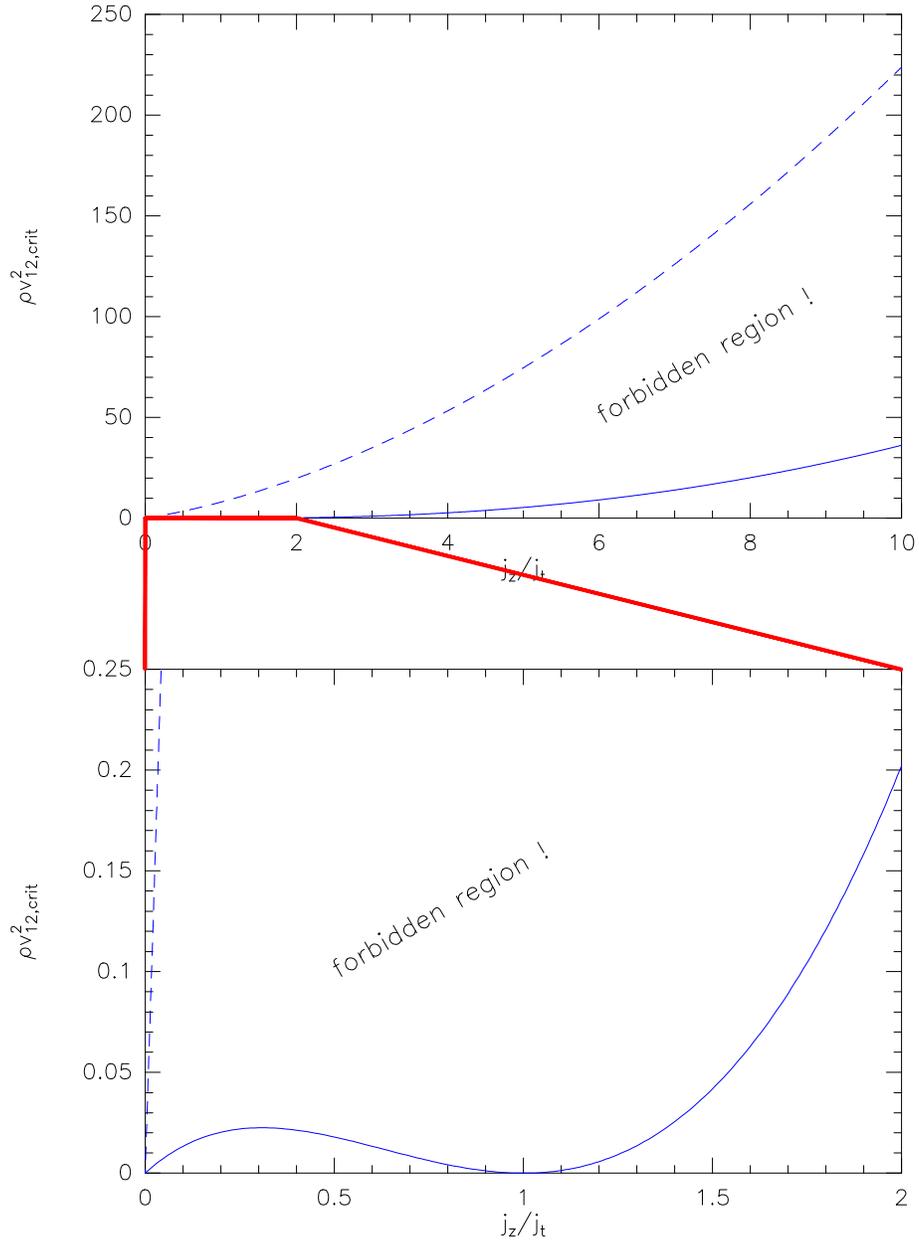}}
\caption{Not all values for the \lq free\rq~velocity
parameter $V_{12}$ are allowed. There exist some forbidden regions
for the velocity parameter.}
\label{bild}
\end{figure}

For CII it is clear that the criterion is given by
\begin{eqnarray}
\rho V_{12}^2 &\leq & \frac{\mu_{0}}{2}\left( 3 j_{t} + j_{z}\right)
\gamma j_{z} - 2 \mu_{0}\gamma |j_{z}| \sqrt{\frac{ j_{t}
\left( j_{t} + j_{z}\right)}{2}} 
\, .\label{35}\end{eqnarray}
For $j_{z}<0$ no value of $V_{12}$ can satisfy the relation
Eq.\,(\ref{35}). Only for $j_{z}\geq 0$ solutions exist in the region beyond
the curve Fig.\,\ref{bild} and the $j_{z}$-axis.

Thus we can summarize: For $j_{z}\geq 0$ the allowed values for
$\rho V_{12}$ are given by the relation
\begin{eqnarray}
\epsilon_{2}\geq\rho V_{12}^2\geq 0\quad\lor\quad \rho V_{12}^2\geq\epsilon_{1}\, .
\end{eqnarray}

 \subsubsection{Solutions with shear flows}

In the case of a vanishing $V_{12}$ or $V_{21}$ the equations again lead to the statement
that either $V_{12}=V_{21}=0$ in both cases or that again $\gamma=0$. This implies
that the allowed flows are of higher orders with respect to the spatial variables,
like indicated in \inlinecite{priest75}$\!\! ,$ or need special, maybe unphysical
thermodynamical constraints.
Also the case of general shear flows leads to $\gamma=0$.
Such solutions will not be discussed here.

\subsubsection{The resistivity}

The resistivity is given by the $z$-component of
$\vec{\rm v}\times\vec B$ and results in the quadric
\begin{eqnarray}
\eta=\frac{1}{j_{z}}\left(2aV_{12}y_{0}x-\left(2a V_{12}+2b V_{21}\right)xy\right)\, .
\end{eqnarray}

Neglecting the term with $y_{0}$,
the above equation shows that the isocontours of the resistivity have four
branches. In two quadrants $\eta$ has a positive sign and the other two quadrants
represent an effective negative resistivity.

This example of unbounded fields shows an interesting behaviour because
the solutions are exact analytical ones and they exhibit the fact that there is
a special relation between the derivatives of $\vec{\rm v}$ and of $\vec B$
connected about algebraic varieties which generate the solutions. We stop these
investigations here due to the problem of negative 
resistivity. To interpret this negative effective resistivity
(or negative non-idealness) and 
search for a reasonable physical interpretation or correponding
physical mechanism needs further investigation.

\subsection{Complete linearization of the resistive system concerning $x$ and $y$}
\label{coli}
Here $V_{11}\neq 0$ in general, as the equation which contradicts the
thermodynamical constraints, the last two equations of
Eqs.\,(\ref{18}) and \,(\ref{19}), being part of the energy equation,
are neglected due to their higher order in $x,y$.

In this case we only need to regard equations, being of first order in $x$ and
$y$, namely
\begin{eqnarray}
 \rho\left(V_{11}^2 + V_{12} V_{21}\right) + 2 p_{4} - 2 a j_{z} &=& 0
\, ,\label{cl9}\\
 \rho\left(V_{11}^2 + V_{12} V_{21}\right) + 2 p_{5} - 2 b j_{z} &=& 0
\, , \label{cl10}\\
 p_{3} &=& 0\, , \label{cl11}\\
 - \rho( V_{12} V_{21} +
V_{11}^2) y_{0} + p_{2} +\rho g &=& 0\, ,\label{cl12} \\
-\rho V_{11}^2 x_{0}-\rho V_{12} V_{21} x_{0} + p_{1} &=& 0\, ,
\label{cl13}\end{eqnarray}

\begin{eqnarray}
 (-V_{11} p_{1} -V_{21} p_{2})x_{0} + (V_{11} p_{2} -V_{12} p_{1})y_{0} &=&
0\, , \label{cl3}\\
 V_{21} p_{2} +V_{11} \left( p_{1} -2 x_{0} p_{4}\right) - 2V_{12}y_{0} p_{4} &=&
2aq( V_{11} x_{0} + V_{12}y_{0})  \, ,\\
 V_{12} p_{1} - 2 V_{21} x_{0} p_{5} + V_{11}\left( 2  y_{0} p_{5} - p_{2}\right) &=&
2bq\left(  V_{21} x_{0} - V_{11} y_{0}\right) \, , \label{cl4}
\end{eqnarray}
which contain only terms of first order.
This is a nonlinear system consisting of eight, but, of course,
effectively seven equations for nine unknowns.

\subsection{Solutions without gravity and shift}\label{sws}

\begin{figure}[t!]
\resizebox{\hsize}{!}{\includegraphics{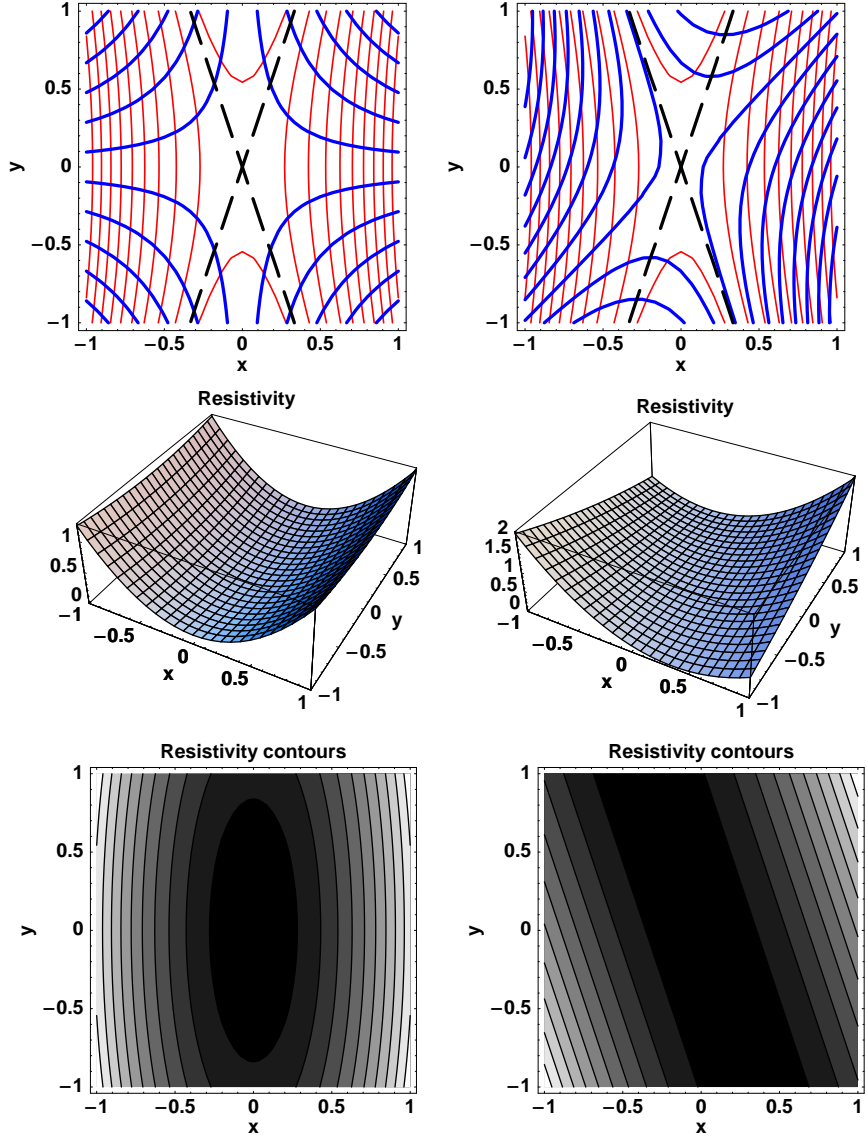}}
\caption{Flow for the parameters $V_{11}=V_{12}=1$ and $s=0$ (left column)
and $s=-1$ (right column), crossing all four magnetic separatrix branches
(top panels), shape of the positive resistivity (middle panels), and 
the elliptic shaped contours of the resistivity (bottom panels).}
\label{cross4}
\end{figure}

In this case the stagnation point is identical with the magnetic null
point. The general solutions is now represented by
\begin{eqnarray}
p_{4} &=& -\frac{\rho}{2}\left(V_{11}^2+V_{12}V_{21}\right)+ aj_{z}\nonumber\\ 
p_{5}&=& -\frac{\rho}{2}\left(V_{11}^2+V_{12}V_{21}\right)+ b j_{z} \, ,
\label{cl15}
\end{eqnarray}
with free parameters $V_{11}$ and $V_{12}$ (and, of course,
the \lq magnetic parameters\rq~the threshold current $j_{t}$ and the
actual current $j_{z}$).
The parameter $V_{21}$ is not free, as it is determined by the fact that the
quadric (surface), representing the resistivity, must be an elliptic paraboloid.
We have to introduce another parameter $s$ to express $V_{21}$ as function of
$V_{12}$ and $s$.
The elliptic paraboloid is the only quadric that allows a positive resistivity
with one zero (the vertex of the paraboloid) at the magnetic null point. 
All other quadrics can be excluded, with the exception of the degenerated case
of a cylindrical paraboloid, see, e.g.,\inlinecite{bronstein} or
\inlinecite{bartsch}.
The cylindrical paraboloids are limiting cases with $s=\pm 1$ as is shown
as in the example in Fig.\,\ref{cross4} (here the specific value is $s=-1$).
The general solution is then parameterized by $V_{11}, V_{12}, s$
\begin{eqnarray}
V_{21}=\frac{j_{t}+j_{z}}{j_{t}-j_{z}} V_{12}+2s V_{11}\,
\frac{\sqrt{j_{t}^2-j_{z}^2}}{j_{t}-j_{z}}\, ,
\label{cl16}\end{eqnarray} 
with the restriction $s\in[-1,1]$.
Multiplication of Eq.\,(\ref{cl16}) with $V_{12}$,
inserting this expression into the expression for the Eigenvalue
$\lambda_{V}^2 = V_{11}^2+V_{12} V_{21}$, and completing the square gives
\begin{eqnarray} 
\lambda_{V}^2 &=& V_{11}^2+V_{12} V_{21}\nonumber\\
&=&\left(V_{11} + s V_{12}\frac{\sqrt{j_{t}^2-j_{z}^2}}{j_{t}-j_{z}}
\right)^2 + \left(1-s^2\right) V_{12}^2\frac{j_{t}+j_{z}}{j_{t}-j_{z}}\geq 0 \, .
\end{eqnarray}
If $|j_{z}|<j_{t}$ and $|s|<1$ as assumed, the magnetic field is of hyperbolic
type and it is guaranteed that the stagnation point is also of hyperbolic type,
i.e. has two purely real Eigenvalues, namely one positive value and a negative
counterpart.

Although the solutions shown here are no reconnection solutions, as the electric
field has to be zero, because only resistive dissipation is included
in our investigation, they represent almost reconnection solutions(=reconnection-like
solutions) where the
necessary condition for reconnection is fulfilled.

Let us investigate now the geometry, respectively the slopes of
the separatrices of the flow and the magnetic field.
Defining $J_{z}:=j_{z}/j_{t}$, the magnetic separatrix is given by
$A=0$ and
$A=\frac{\mu_{0}}{4}\left[ \left(j_{t}-j_{z})y^2 -\left(j_{t}+j_{z}
\right)x^2\right)\right]$ (we restrict this to $|J_{z}|<1$)
and can therefore be expressed by
\begin{eqnarray}
\sqrt{1-J_{z}} \, y\pm \sqrt{1+J_{z}} \, x = 0\,\,\Leftrightarrow\,\,
y=\pm\frac{\sqrt{1+J_{z}}}{\sqrt{1-J_{z}}}x:=\pm K_{m} x \, ,
\end{eqnarray}
where $\pm K_{m}$ is the slope of the both magnetic separatrix lines or to say
the both asymptotes.
As $V_{21}$ is given by the general solution, Eq.\,(\ref{cl16}), by defining a 
stream function $\psi$ via
\begin{eqnarray}
{\rm v}_{x}=\frac{\partial\psi}{\partial y}\quad\textrm{and}\quad
{\rm v}_{y}=-\frac{\partial\psi}{\partial x}\, ,
\end{eqnarray}
we can integrate the above equations and get for the stream function
\begin{eqnarray}
\psi=\frac{V_{12}^2}{2} y^2 - \frac{V_{21}^2}{2} x^2 + V_{11} x y \, .
\end{eqnarray}
The fluid separatrix is here given by $\psi=0$, geometrically this
are asymtotical lines (asymptotes).

We will briefly discuss the problem that the magnetic separatrix is partially
identical with the hydrodynamic separatrix. In this case the plasma flow can only
take place across one part of the separatrix, or both separatrix lines are
identical and no reconnection can take place.


One can clearly recognize that in the case $V_{11}=0$ and thus $(V_{21}/V_{12})^2
=(1+J_{z})/(1-J_{z})$ both corresponding asymtotical branches($=$separatrix lines)
have the same slope. Thus the hydrodynamical separatrix and the magnetic
separatrix are identical, the plasma cannot cross the magnetic separatrix and
therefore no reconnection can take place.

Writing $y(x)=K_{V}x$ (if $V_{11},V_{12}\neq 0$), 
where $K_{V}$ is the slope of the hydrodynamic separatrix, and inserting this into
$\frac{V_{12}^2}{2} y^2 - \frac{V_{21}^2}{2} x^2 + V_{11} x y=0$
with the parametric expression for $V_{21}$ in Eq.\,(\ref{cl16}), we get the slopes
$K_{V1,V2}$ of the two asymptotes/separatrix lines
\begin{eqnarray}
K_{V1,2} &&=-\frac{V_{11}}{V_{12}}\pm\sqrt{\frac{V_{11}^2+ V_{12} V_{21}}{V_{12}^2}}
\nonumber\\
 &&=\frac{-V_{11} \pm \sqrt{V_{11}^2+2sV_{11}V_{12}
\frac{\sqrt{1-J_{z}^2}}{1-J_{z}} + V_{12}^2\,\frac{1+J_{z}}{1-J_{z}}}}{V_{12}}
\nonumber\\ \nonumber\\
 &&=-V_{rel} \pm \sqrt{V_{rel}^2+2sV_{rel}
\frac{\sqrt{1-J_{z}^2}}{1-J_{z}} + \frac{1+J_{z}}{1-J_{z}}}\,\,\,\, ,
\label{KV}
\end{eqnarray}
where $V_{rel}=V_{11}/V_{12}$. For $V_{12}=0\neq V_{11}$ the asymptotes are given by $x=0$ and
$y= s\sqrt{(1+J_{z})/(1-J_{z})}x$,
and for $V_{12}=V_{21}=0$ we get $x=0$ and $y=0$ as separatrices.

Let $V_{11}=V_{12}(=1)$, then
\begin{eqnarray}
K_{V1,2} = -1 \pm \sqrt{\left(1+s\frac{\sqrt{1-J_{z}^2}}{1-J_{z}}\right)^2}
= -1 \pm \left(1 + s\frac{\sqrt{1-J_{z}^2}}{1-J_{z}}\right)\,\, .
\end{eqnarray}
There is {\it one} corresponding slope of the flow separatrix
with respect to one of the magnetic separatrix lines, if $s=\pm 1$.

Therefore we can formulate the following theorem (in analogy to
known anti-reconnection theorems), which is now restricted
to almost reconnective solutions with vanishingly small electric fields and
resistive disspation only, i.e. vanishing non-resistive dissipation
mechanisms:

\paragraph*{Theorem} {\it If the flow close to the null point is in good approximation
incompressible, the electric field is negligible and other kinds of
dissipation mechanisms than the resistive dissipation are also
negligible, then the plasma flow cannot cross the magnetic separatrices
if for the both slopes of the both magnetic separatrix lines
$K_{m}=K_{m1}=-K_{m2}$ with $K_{m}=\sqrt{(1+J_{z})/(1-J_{z})}$
and the following is valid: 
\begin{eqnarray}
&& I.\,\, a)\, K_{m}=K_{V1}\quad\land\quad b)\, -K_{m}=K_{V2}\nonumber\\
&&\textrm{or}\nonumber\\
&& II.\,\, a)\, K_{m}=K_{V2}\quad\land\quad b)\, -K_{m}=K_{V1}
\end{eqnarray}
If only I.a) or I.b) or II.a) or
II.b) is valid then only one magnetic separatrix can be crossed and
as the other stream lines converge to the second magnetic separatrix line
without crossings, one could call this reconnective annihilation,
see, e.g., Priest \& Forbes (2000). The necessary condition for a complete
non-crossing is $V_{11}=0$ for $V_{12}\neq 0$
and for only partly crossing flows $s=\pm 1$.}

For almost all values of $s$, here $s=0$, the flow in Fig.\ref{cross4} (left column)
crosses all four separatrix branches (all two separatrix lines),
the resistivity is positive (only zero at the null point, middle
panel), and their isocontours are ellipses (bottom panel).
For the special parameter $s=-1$ it can be seen in Fig.\,\ref{cross4} (right column) that
the flow, like for the aforementioned magnetic reconnective annihilation
solutions, crosses only the separatrix line with the positive slope,
while it converges to the other magnetic separatrix, being almost field aligned
(top panel), the resistivity being also postive (middle panel), and the
isocontours of the resistivity are straight lines (bottom panel).

We now focus on the effective resistivity and therefore on the nature of
ohmic heating/disspation.
The resistivity is given by
\begin{eqnarray}
\eta=\frac{\mu_{0}}{2}\frac{V_{11}}{j_{z}}\left[ \left(j_{t}+j_{z}\right) x^2
-2s\sqrt{j_{t}^2-j_{z}^2} xy + \left(j_{t}-j_{z}\right) y^2\right] .
\label{resi1}
\end{eqnarray}
To ensure that the dissipation is positive we prove that the term
$\left(j_{t}+j_{z}\right) x^2
-2s\sqrt{j_{t}^2-j_{z}^2} xy + \left(j_{t}-j_{z}\right)y^2$
is larger than zero (or zero) for all $s$ and $x,y$.
Let $K_{1,2}=j_{t} \pm j_{z}$ and $K_{1}, K_{2}>0$ as required for an X-point.
Then
\begin{eqnarray}
0 &&\leq \left(\sqrt{K_{1}}x-s\sqrt{K_{2}}y\right)^2=K_{1}x^2-2s\sqrt{K_{1}K_{2}}xy
+s^2 K_{2}y^2\nonumber\\
&&\leq K_{1}x^2-2s\sqrt{K_{1}K_{2}}xy+ K_{2}y^2. \quad\square
\end{eqnarray} 
This implies that the sign of the resistivity $\eta$ depends on the sign of
$\frac{V_{11}}{j_{z}}$. As the dissipation should be positive, i.e. $\eta>0$,
thus $V_{11}$ and $j_{z}$ must have the same sign.

For $s=\pm 1$ it can be recognized from Eq.\,(\ref{resi1}) and from
the bottom panel Fig.\,\ref{cross4} (for the case $s=-1$)
that the resistivity contour lines are straight lines given
by $\eta=\mu_{0} V_{11}/(2j_{z})\,(\sqrt{j_{t}+j_{z}}x\pm \sqrt{j_{t}-j_{z}} y)^2$,
and therefore the resistivity has the structure of a one-dimensional sheet.
The case $s=\pm 1$ therefore reflects the 1D character of the non-idealness,
which can also be found in the papers of \inlinecite{craighenton}.

\section{Conclusions}

We analyse the solution space of the MHD system with a generalized non-idealness in
Ohm's law close to an X-type magnetic null point. This procedure is done
to get non-ideal/resistive but
formally non-reconnective solutions that can be regarded as reconnection
solutions with a vanishing electric field. Such solutions show the parameters
and their relation amongst each others, to get non-ideal non-reconnective, i.e. either
almost reconnective annihilation or completely almost reconnective (= reconnection-like)
solutions.

We assume that the flow close
to the magnetic null point is governed strictly by the pure non-ideal/resistive system of
MHD equations, and that the stagnation point is close to the null
point (or that both singular points in the plane are identical).
The focus is on the region close to the magnetic null point to investigate
the possible type of topology of the corresponding stagnation point (flow).
To use a closure for the equations and to analyse the resistive MHD system
we use the corresponding energy equation.

We further assume that the density is constant close to the stagnation point. This
assumption is not a necessary, but a plausible condition concerning the mass
distribution in the vicinity of stagnation points. This leads to a vanishing
divergence of the plasma velocity and as a logical consequence to a vanishing
resistivity close to the stagnation point. Inserting this into resistive Ohm's
law implies that the electric field has to vanish at the stagnation point.
We  conclude that the electric field has to be zero, and as a consequence of
the 2D stationary approach the electric field has to be zero everywhere.

This case of vanishing divergence of the velocity field and thus of a vanishing
electric field can be regarded as a physical approximation or limit. Obviously
a classical reconnection process requires a non-vanishing divergence of the
velocity field to produce
a non-zero reconnection rate, determined by the constant electric field $E_{0}$,
see, e.g.,\inlinecite{priestforbes}.

From resistive Ohm's law one can clearly
recognize that close to the magnetic null point the electric field is identical to
the \lq convective\rq~electric field generated by the product from resistivity and
current density. As the electric field is zero, the resistivity inside the finite
current sheet must be zero at the magnetic neutral point. The effect is that
the
resistivity cannot be constant close to the magnetic neutral point. Therefore
the resistive Ohm's law is an equation determining the topological
type of the magnetic field and the velocity field and connecting/relating
it to the shape of the resistivity, which is a quadratic function of the spatial
coordinates.
We call this spatially varying resistivity {\it effective resistivity} or
for short resistivity as usual.
As the current density is constant close to the magnetic null point, the
right hand side of Ohm's law, Eq.\,(\ref{6}) could be regarded as a
generalized non-ideal term.

A {\it necessary} condition for reconnection is that the plasma flow can cross
magnetic separatrices.
We found that crossing of magnetic separatrices requires also an X-type stagnation point
flow if the effective resistivity and thus the dissipation should be positive.

The sufficient criterion for reconnection
is the crossing of the separatrix and the non-vanishing electric
field. This reconnection process can take place for/in the incompressible case if
there is not only a resistive, but an additional 
non-resistive dissipation (term) that occurs only in the energy equation.
Therefore one possibility could be heat conduction.
Another possibility is the viscous case with constant viscosity. Here, the
Laplacian of the velocity field vanishes in the first order momentum equation
of the ions (Navier-Stokes) as the velocity field is linear. 
These cases have to be investigated in more detail but
have been discussed under the assumption of constant resistivity, e.g., by 
\inlinecite{priestforbes}.

The analyses started by us here will be extended to configurations that have
density gradients close to the stagnation point and therefore allow for a non-zero
electric field.
The extension to energy equations allowing for a deviation from the classical
energy equation of resistive MHD will provide us with a relation between
the deviation from pure resistive dissipation and deviation from the classical
(X-type) stagnation point flow.

\begin{acks}
D.H. Nickeler acknowledges financial support from GAAV\,\v{C}R grant No. IAA300030804
and M. Karlick\'{y} from GA\,\v{C}R grant No. 300030701. D.H. Nickeler is grateful to Dr. Michaela
Kraus for the help in preparing Fig.2 and careful reading of the manuscript.
\end{acks}

\end{article}

\end{document}